\documentclass[12pt]{article}
\usepackage{amssymb,graphicx,times}
\usepackage{amssymb,amsmath,amsfonts,amsthm,epsf,times,graphicx}
\usepackage{delarray,psfig,psfrag}
\newcommand{\eps}{\varepsilon}
\newcommand{\fd}{f_{d}}

\newcommand{\mud}{\mu_{d}}
\newcommand{\sigmad}{\sigma_{d}}
\newcommand{\gammad}{\gamma_{d}}

\newcommand{\ta}{t_{1/2}}
\newcommand{\tb}{t_{3/8}}
\newcommand{\tc}{t_{1/4}}

\newcommand{\za}{z_{1/4}}
\newcommand{\zb}{z_{1/8}}

\newcommand{\ya}{y_{1/8}}

\def\ac{\mathcal{A}}

\newcommand{\flu}[1]{\delta #1}

\newcommand{\real}{\mathbb{R}}

\newcommand{\beq}{\begin{eqnarray}}
\newcommand{\eeq}{\end{eqnarray}}
\newcommand{\bea}{\begin{eqnarray}}
\newcommand{\eea}{\nonumber \end{eqnarray}}
\newcommand{\nn}{\nonumber\\}

\newcommand{\eq}[1]{(\ref{#1})}
\theoremstyle{plain} 
\newtheorem{theorem}{Theorem}[section]  
\newtheorem{proposition}{Proposition}[section]
\newtheorem{lemma}{Lemma}[section]
\newtheorem{corollary}{Corollary}[section]
\newtheorem*{notation}{Notation}        
\newtheorem{definition}{Definition}[section]
\theoremstyle{definition}
\newtheorem{example}{Example}[section]

\newtheorem{remark}{Remark}[section]

\theoremstyle{remark}

\newcommand{\bthm}{\begin{theorem}}  
\newcommand{\ethm}{\end{theorem}}
\newcommand{\bpro}{\begin{proposition}}
\newcommand{\epro}{\end{proposition}}
\newcommand{\bdefn}{\begin{definition}}
\newcommand{\edefn}{\end{definition}}
\newcommand{\bcor}{\begin{corollary}}
\newcommand{\ecor}{\end{corollary}}
\newcommand{\blem}{\begin{lemma}}
\newcommand{\elem}{\end{lemma}}
\newcommand{\bexp}{\begin{example}}
\newcommand{\eexp}{\end{example}}
\newcommand{\bnotn}{\begin{notation}}
\newcommand{\enotn}{\end{notation}}
\newcommand{\brem}{\begin{remark}}
\newcommand{\erem}{\end{remark}}
\newcommand{\bprf}{\begin{proof}[{\sc Proof.}] $\,$} 
\newcommand{\eprf}{\end{proof}}   

\newcommand{\rbr}[1]{\left(#1\right)}
\newcommand{\sbr}[1]{\left[#1\right]}
\newcommand{\cbr}[1]{\left\{#1\right\}}
\newcommand{\mo}[1]{\left|#1\right|}

\newcommand{\eval}[2]{\left .#1\right|_{#2}}

\newcommand{\de}{\partial} 
\newcommand{\der}[2]{\frac{d #1}{d #2}} 
\newcommand{\pder}[2]{\frac{\partial #1}{\partial #2}} 

\newcommand{\bmat}{\begin{bmatrix}}
\newcommand{\emat}{\end{bmatrix}}
\newcommand{\brmat}{\begin{pmatrix}}
\newcommand{\ermat}{\end{pmatrix}}

\begin{document}

\vspace*{0.88truein}
\centerline{\bf OPTIMAL HEDGING OF DERIVATIVES WITH}
\vspace*{0.087truein}
\centerline{\bf TRANSACTION COSTS}
\vspace*{0.087truein}
\vspace*{0.37truein}
\baselineskip=10pt
\centerline{\footnotesize ERIK AURELL}
\vspace*{0.015truein}
\centerline{\footnotesize\it AlbaNova University Center}
\baselineskip=10pt
\centerline{\footnotesize\it Department of Physics}
\baselineskip=10pt
\centerline{\footnotesize\it KTH - Royal Institute of Technology}
\baselineskip=10pt
\centerline{\footnotesize\it SE-106 91 Stockholm, Sweden}
\baselineskip=10pt
\centerline{\footnotesize\it erik.aurell@physics.kth.se}
\vspace*{0.37truein}
\centerline{\footnotesize PAOLO MURATORE-GINANNESCHI}
\vspace*{0.015truein}
\centerline{\footnotesize\it Departments of Mathematics and Statistics}
\baselineskip=10pt
\centerline{\footnotesize\it University of Helsinki PL 68}
\baselineskip=10pt
\centerline{\footnotesize\it FIN-00014 Helsingin Yliopisto Finland}
\baselineskip=10pt
\centerline{\footnotesize\it paolo.muratore-ginanneschi@helsinki.fi}
\baselineskip=10pt
\vspace*{10pt}
\begin{abstract}
We investigate the optimal strategy over a finite
time horizon for a 
portfolio of stock and bond and a derivative
in an multiplicative Markovian market model with transaction costs (friction). 
The optimization problem is solved by 
a Hamilton-Jacobi-Bellman equation, which
by the {\it verification theorem} 
has well-behaved solutions if certain conditions 
on a potential 
are satisfied. 
In the case at hand, these conditions simply imply  
arbitrage-free (``Black-Scholes'') pricing of the derivative.
While pricing is hence not changed by friction
allow a portfolio to fluctuate around a delta hedge. 
In the limit of weak friction, we 
determine the optimal control to essentially be of two parts: 
a strong control, which tries to bring the stock-and-derivative
portfolio towards a Black-Scholes delta hedge; and 
a weak control, which moves the portfolio by adding
or subtracting a Black-Scholes hedge.
For simplicity we assume growth-optimal investment criteria
and quadratic friction.
\end{abstract}

\section{Introduction}  
\label{s:introduction}
\vspace*{-0.5pt}
\noindent
An idealised model of investment is a sequence of gambles where an investor
at each time step decides if to re-balance her investments, and, 
if so, by how much. 
The game is multiplicative if the pay-off is proportional to capital, 
and Markov if the new capital and new position only depend on the previous 
state and the action taken then.
In two previous contributions \cite{AurellMG1,AurellMG2} we computed
the strategy an investor should use to maximize the growth
rate of her wealth, in the presence of transaction costs,
if she can invest in stock and bonds.
In this paper we extend the investment possibilities to also include
a derivative security, {\it e.g.} an option on the stock.

Asset allocation optimization in the presence of transaction costs has a 
long and distinguished history in finance. The main mathematical tool is the 
Hamilton-Jacobi-Bellman equation, introduced in the friction-less case 
by Merton~\cite{Merton68,Merton}, and with friction by 
Constantinides~\cite{Constantinides}. A pedagogical introduction to the
application of the Hamilton-Jacobi-Bellman equation to financial problems 
can be found in \cite{Bjork}.
Our first paper, \cite{AurellMG1}, overlaps with the subsequent work
of Atkinson, Pliska and Wilmott~\cite{AtkinsonPliskaWilmott},
where also a perturbative expansion around the friction-less limit
is performed. In \cite{AurellMG2} we treated also optimization 
over a finite-time horizon.

Turning to option pricing and hedging in the presence of transaction
costs, in several contributions, the Hamilton-Jacobi-Bellman approach was not used. 
A classical paper in this line of research, leading to a modified option price,
is that of Leland~\cite{Leland}, another one, also leading to a modified
option price, is that of Bouchaud and Sornette~\cite{BouchaudSornette}.
Boyle and Vorst~\cite{BoyleVorst} looked at a portfolio replicating 
an option with transaction costs. This is a different setting to that
of optimal control and an Hamilton-Jacobi-Bellman  approach.
Avellaneda and Paras~\cite{AvellanedaParas} did consider Hamilton-Jacobi-Bellman , 
but in the limit of large transaction costs.
Other notable contributions are those of Edhirisinghe, Naik and Uppal 
\cite{EdhirisingheNaikUppal}, Davis, Panas and Zariphopolou 
\cite{DavisPanasZariphopolou}, 
and Bensaid, Lesne, Pages and Scheinkman
\cite{BensaidLesnePagesScheinkman}.

A recent contribution considering a portfolio including stock
and derivatives in the Hamilton-Jacobi-Bellman  approach is the paper of
Constantinides and Zariphopolou \cite{ConstantinidesZariphopolou}.
However, these authors consider derivatives which 
can be traded only once (section 3 of that paper), while we
address the situation where derivatives are traded continuously.
Constantinides and Zariphopolou show that 
expected utility is increased by including derivatives, and 
that derivative prices must obey certain bounds in their model. 
We note that Constantinides and 
Zariphopolou~\cite{ConstantinidesZariphopolou}
remark that the bounds they derive 
would be tighter if derivatives would be traded more than once.

In our treatment, the Hamilton-Jacobi-Bellman  equation for the stock, 
bond and derivative problem leads to time-dependent controls, as in the 
simpler case of~\cite{AurellMG2}.
A feature which appears when stock and derivative are both traded continuously
is that the optimization problem in general ill-defined, unless the 
price process obeys a solvability criterion, known as the {\it verification theorem} 
in the mathematical theory of controlled stochastic processes 
\cite{FlemingMeteSoner}. 
In Hamilton-Jacobi-Bellman  language, the value function is
potentially unbounded, because the number of variables that can be controlled
(the positions in stock and derivative) is larger than the number of independent 
noise sources. In the case at hand, the potential is the expected utility of 
the portfolio as a function of the fraction of wealth invested in stock and 
derivative, and the conditions simply imply  arbitrage-free (``Black-Scholes'') 
pricing of the derivative.

While we hence find that pricing does {\it not} depend on market friction,
the optimal investment strategy does.
Qualitatively speaking, we determine the optimal control to be of two parts: 
a strong control, which tries to bring the stock-and-derivative
portfolio close to a Black-Scholes delta hedge; and a weak control, which moves 
the portfolio by adding or subtracting a Black-Scholes hedge.
The rationale for the presence of the weak control is that the strong control 
acts to oppose the underlying diffusion of the portfolio, in the direction
normal to the delta hedge. The larger that diffusion, the higher will be the 
average friction costs, per unit time. It is therefore advantageous to invest 
as much in the delta hedge to make the diffusion in the normal direction as
small as possible.  
As in~\cite{ConstantinidesZariphopolou} we find that 
introducing derivatives increases expected utility. Essentially, in the small 
friction limit, expected transaction costs decrease, and dependence upon friction 
parameter is pushed to higher order.

A technical contribution in this paper, analogous to
\cite{AtkinsonPliskaWilmott,AurellMG2} (without derivatives), 
is that we introduce 
a multi-scale expansion around the friction-less limit.
Since, however, we have two independent variables under control (the stock and the
derivative), we can have different scales in two different
directions. In fact, we will show that in the weak-noise limit
there is a fast control direction, and a slow control direction. 
The fast control strives to bring the portfolio to an optimal stock
portfolio plus a Black-Scholes hedge. Financially, this means that
an optimal investment strategy is to hold some amount in stock, and then 
some number of fully hedged derivatives. That number is however controlled 
on a longer time scale, by the slow control.
Two limit cases are of interest. First, far from expiry a derivative
is not much different from stock, and the situation is similar to
only investing in stock and bonds.
Second we can also deal with the situation
close to expiry. There, the best strategy turns out to be to
hold little funds proportional to the Black-Scholes hedge, 
i.e. to avoid derivatives in the optimal strategy. This concurs
with the practice of closing out positions in derivatives before
expiry. 

For simplicity we work in this paper with quadratic friction. 
These can be motivated as an effective description of market impact 
(see {\it e.g.}~\cite{Farmer}).
The reader is referred to \cite{AurellMG2} for details.
Linear friction costs, arguably more realistic, 
lead to free boundary problems in the Hamilton-Jacobi-Bellman formalism, 
which are considerably harder, from the analytical and numerical point of view.

For simplicity we will also furthermore assume throughout that an investor
strives to optimize expected growth of capital, which in a multiplicative 
market model means logarithmic utility. Growth optimal strategies
were first introduced by Kelly in the context of information theory 
\cite{Kelly}. 
Growth-optimal strategies have the well-known property of eventually,
for long times, outperforming any other strategy with probability 
one~(\cite{HakansonZiemba} and references therein), 
but do not maximize vanilla-flavored utility functions, 
see {\it e.g.}~\cite{DybvigRogersBack}.
In the present context, logarithmic utilities should merely 
be looked upon a definite and convenient choice, which brings some
mathematical tidiness. 

The paper is organised as follows. In section  \ref{sec:model} 
we state the model (without derivative), and the controls we consider.
We state the optimisation
problem in the framework of the Hamilton-Jacobi-Bellman equation.
In section \ref{sec:HJB} we show that the non-linear Hamilton-Jacobi-Bellman 
equation governing the dynamics, in our example,
is solvable in the small transaction costs limit by means of a multi-scale 
perturbation theory (see for example \cite{Bocquet}, or \cite{Frisch}, 
chapter 9). This is the main technical result of the paper, and reduces the 
non-linearity to a normal form. 
All higher order corrections can be computed from ancillary linear non-homogeneous 
equations. In section \ref{sec:quantitative} we solve analytically the normal form 
of the non-linearity. 
The approximation turns out to be very accurate for realistic values of the 
parameters in the model. The last section is devoted to a discussion of the results.

\section{Bond, stock and derivatives}
\label{sec:model}

In this section we define notation, and state the problem.
The market consists of a risk-less security (bond, or bank account)
and a risky security (stock).
By a change of numeraire we take the price of
the risk-free security to be constant in time.
The stock price is taken the standard 
log-normal process:
\begin{equation}
\psi_{t+dt} = \psi_t\left(1+ \mu dt + \sigma dB_{t}\right)
\label{model:price}
\end{equation}
Here $dB_{t}$ denotes the stochastic differential and 
$\mu$ and $\sigma$ are positive constants.
Nothing in the following analysis would essentially
change if $\mu$ and $\sigma$ would be functions of $t$ and $S_t$, 
as long as the market is still complete, see {\it e.g.}
\cite{KaratzasShreve}. Consider now first a portfolio in only stock and bond.
The control variable then is the fraction of wealth an investor has
invested in stock:
\begin{equation}
\rho_{t}  = \frac{ W_{t}^{\mathrm{Stocks}}}{W_{t}}
\label{model:firstinvestment}
\end{equation}
This variable changes both in result to market fluctuation, 
i.e.~\eq{model:price}, and re-hedging. We assume that a control 
can be executed of the form
\begin{equation}
d\rho^{\mathrm{control}}_t = f dt
\label{model:friction} 
\end{equation}
and doing so carries a cost
\begin{equation}
dW_t^{\mathrm{trading costs}} = -\gamma F(f) W_{t} dt
\label{model:continuumfriction}
\end{equation}
where $F$ is a semi-positive definite functional of the stochastic control.
The form of $F$ models the transaction costs.

The coupled stochastic differential equations of $W$ and $\rho$ are then
\begin{eqnarray} 
dW_{t} &=& W\sbr{\rbr{\mu \rho_{t} dt+\sigma \rho_t \,dB_t -\gamma F(f) dt}}
\label{model:wealthsde}\\
d\rho_t &=&\sbr{ f+ \rho_t (1-\rho_t) (\mu-\sigma^2 \rho_t)}\,dt+
\sbr{\sigma \rho_t (1-\rho_t)}\,dB_t+\gamma\,\rho_t F(f)dt
\label{model:fractionsde}
\end{eqnarray}
For a derivation of these equations, see~\cite{AurellMG2}.
The time-dependent growth-optimization problem,
of a stock and bond
portfolio, from time $t$ to some final time $T$,
is simply to choose the control $f$ such that the 
expected value of $\log \frac{W_{T}}{W_t}$ is maximized.
By a change of variable, this is equivalent to maximizing
the expectation value, over the controlled diffusion process,
of a potential (utility function) depending on $\rho$ and $f$.
Implicitly, we assume unbounded borrowing at the risk-less rate, and 
no restrictions on going short. These are not in fact serious
limitations, because the optimal solution, with transaction costs,
is to hold the fraction invested in stock close the optimal
value of $\frac{\mu}{\sigma^2}$, see~\cite{AurellMG1}, which is finite. 
In the main body of this paper, we will use
quadratic friction costs, {\it i.e.} $F(f) = f^2$,
for a discussion of linear friction costs, see
\cite{AurellMG1}.

Let us define a derivative security as a third investment
possibility, the price of which, $C(\psi_t,t)$, only depends on 
the moment of time $t$ and the price of stock. 
The price dynamics of the derivative is
\begin{eqnarray}
\frac{d C}{C}= 
\frac{1}{C}\sbr{\de_t C+\mu \psi_t \de_{\psi_t}C+\frac{\sigma^2\,\psi_t^2}{2}
\de_{\psi_t}^2 C}dt+ 
\frac{\sigma \psi_t \de_{\psi_t}C}{C}\, dB_t:=\mu_d dt+\sigma_d\,dB_t
\end{eqnarray}
where we for later convenience introduce amplitudes $\mud$ and $\sigmad$. 
Both are of course functions of $t$ and $\psi_t$.
Let now as before the fraction invested in stock be $\rho$
with control $f$, and the fraction invested in derivative $\eta$,
with control $\fd$. Exercising either of the controls in a time interval
$dt$ carries a cost $\mathcal{F}(f,\fd) W_t dt$.

The coupled equations for wealth, $\rho$ and $\eta$ are then
\begin{eqnarray}
dW&=& W\sbr{\rbr{\mu \rho+\mud \eta}dt+
\rbr{\sigma \rho+\sigmad \eta}\,dB_{t}-\mathcal{F}(f,\fd)\,dt}
\label{model:wealth}
\\
d\rho &=&\sbr{f+a+\rho\,\mathcal{F}(f,\fd)}\,dt +b\, dB_{t}
\label{model::stocks}
\\
d\eta &=&\sbr{f_{d}+a_{d}+\eta\,\mathcal{F}(f,\fd)}\,dt+b_{d}\, dB_{t}
\label{model:derivatives}
\end{eqnarray}
where the functions in the drift terms are
\beq
&&a:=\mu\,\rho- \rho\,\sbr{\mu \rho+\mu_{d} \eta-
\rbr{\sigma \rho+\sigma_{d}\eta-\sigma}\rbr{\sigma \rho+\sigma_{d}\eta}}
\label{model:stockdrift}
\\
&&a_{d}=\mu_{d}\,\eta-\eta\,\sbr{\mu \rho+\mu_{d} \eta-
\rbr{\sigma \rho+\sigma_{d}\eta-\sigma_{d}}\rbr{\sigma \rho+\sigma_{d}\eta}}
\label{model:derivativedrift}
\eeq
and the functions in the diffusive terms are 
\beq
&&b=\sigma\,\rho-\rho\,\rbr{\sigma \rho+\sigma_{d}\eta}
\label{model:stockdiffusion}
\\
&&b_{d}=\sigma_{d}\,\eta-\eta\,\rbr{\sigma \rho+\sigma_{d}\eta}
\label{model:derivativediffusion}
\eeq
With analytic transaction costs we have 
\beq
\mathcal{F}(f,\fd)=\gamma\,\mo{f}^{2}+\gamma_d\mo{f_d}^{2}
\eeq
with two friction parameters $\gamma$ and $\gamma_d$.
We now state the problem we want to solve. The expected logarithmic growth 
rate is
\beq
\lambda(x,y,p,t;T):=\hbox{E}\sbr{\log\frac{W_T}{W_t}}_{\rho_t=x;\eta_t=y;\psi_t=p}
\eeq
In consequence the logarithmic growth is the expected value of the utility 
function
\beq
U=\mu\,\rho+\mu_{d}\,\eta-\frac{(\sigma\,\rho+\sigma_{d}\,\eta)^2}{2}
-\mathcal{F}(f,\fd)
\label{model:growthrate}
\eeq
over the probability density $P(x',y',p',t'|x,y,p,t)$ is the probability of 
the process $(\rho_t,\eta_t,\psi_{t})$, to reach point $(x',y',p')$ at 
time $t'$, given it was at $(x,y,p)$ at time $t$:
\beq
\lambda(x,y,p,t;T) =\int_t^T dt' \int  U(z) P(x',y',p',t'|x,y,p,t) \, dx'\,dy'\,dp'
\label{model:loggrowth}
\end{eqnarray}
Note in view of \eq{model:price} the probability density factorizes
to
\bea
P(x',y',p',t'|x,y,p,t)=P_{\rho\,\eta}(x',y',t'|x,y,p,t)P_{\psi}(p',t'|p,t)
\eea
Furthermore, the probability density is in general 
non-autonomous as $\mud$ and $\sigmad$ may depend explicitly upon
the time variable.
The problem is now to find controls $f$ and $f_d$ that maximize
the logarithmic growth.

\section{The verification principle and Black-Scholes}
\label{sec:BS}

It is useful to first discuss the friction-less case.
We will then just reproduce standard elementary results in finance,
but in a formulation convenient for the following discussion.
Without transaction costs, the speculator is free to rehedge continuously.
In such a case the optimisation problem is equivalent to finding the supremum,
at any instance of time, of the instantaneous growth rate 
\beq
V=\mu\,\rho+\mu_{d}\,\eta-\frac{(\sigma\,\rho+\sigma_{d}\,\eta)^2}{2}
\label{BS:growthrate}
\eeq
Equation \eq{BS:growthrate} is a degenerate quadratic functional of 
the fraction in stocks and derivatives. The Hessian of 
\eq{BS:growthrate} 
\beq
\mathsf{H}=
\sbr{\begin{array}{cc} 
-\sigma^2 & -\sigma_{d}\,\sigma
\\
-\sigma_{d}\,\sigma & -\sigma^2_{d}
\end{array}}
\eeq
has a zero eigenvalue along the {\em marginal subspace}
\beq
\sigma\,\rho+\sigma_{d}\,\eta=0
\label{BS:marginal}
\eeq
The second eigenvalue is
negative, $h_{s}=-\rbr{\sigma^2+\sigma^2_{d}}$,
associated to the {\em stable subspace}
\beq
\sigma_{d} \rho-\sigma \eta=0
\label{BS:stable}
\eeq
We now make a change of variables
\beq
\rho \,\hat{e}_{10}
+\eta\,\hat{e}_{01}=\frac{\sigma\,\zeta}{\sqrt{\mo{h_s}}} 
\hat{e}_{s}+\frac{\sigma\,\vartheta}{\sqrt{\mo{h_s}}} \hat{e}_{m}
\eeq
where $(\hat{e}_{10},\hat{e}_{01})$ is the canonical basis of $\real^2$ 
and $(\hat{e}_{m},\hat{e}_{s})$ is an orthonormal basis formed by the unit 
vectors respectively spanning the marginal and stable subspaces of the Hessian
matrix $\mathsf{H}$:
\beq
\hat{e}_{m}:=\frac{1}{\sqrt{\mo{h_s}}}
\sbr{\begin{array}{c} \sigmad\\ - \sigma\end{array}}\,,
\qquad
\hat{e}_{s}:=\frac{1}{\sqrt{\mo{h_s}}}
\sbr{\begin{array}{c} \sigma \\ \sigmad\end{array}}
\eeq
The variable $\zeta$ along the stable eigenspace describes a portfolio
in which the investment in derivatives is weighted by the ratio of the 
volatilities
\beq
\zeta=\rho+\frac{\sigma_d}{\sigma}\eta
\eeq
The utility function reads in these new variables
\beq
V=\mu\,\zeta +\rbr{\mud-\frac{\mu}{\sigma}\sigmad}
\frac{\sigma \rbr{\sigmad\,\zeta-\sigma\,\vartheta}}{\sigma^2+\sigmad^2}
-\frac{\sigma^2\,\zeta^2}{2}
\eeq 
This growth rate is a convex function if and only if the 
second term vanishes. This can happen if either of its two factors
are zero. The first possibility gives the
following {\em solvability condition}:
\beq
\mu-\frac{\sigma}{\sigma_{d}}\mu_{d}=0 \qquad \Rightarrow \qquad
\de_{t^{\prime}}C+\frac{\sigma^2\,\psi^2_{t^{\prime}}}{2}
\de_{\psi_{t^{\prime}}}^2C=0
\label{BS:blackscholes}
\eeq 
holding for every $t^{\prime}\in [t,T]$ and in particular for 
$t^{\prime}$ equal to $t$:
\beq
\de_{t}C+\frac{\sigma^2\,p^2}{2}\de_{p}^2C=0
\label{BS:BS}
\eeq
This is the of course simply Black-Scholes equation at zero interest rate.
The second possibility is that the linear combination 
\beq
\sigmad\,\zeta-\sigma\,\vartheta=0
\eeq
vanishes, which simply means
that the fraction invested in derivatives is zero. 
Optimisation can then be carried out along the stable 
manifold. The utility
\beq
V=\mu\,\zeta-\frac{\sigma^2\,\zeta^2}{2}
\eeq
has a maximum for
\beq
\zeta_{\star}=\frac{\mu}{\sigma^2}
\label{BS:blackscholeshedging}
\eeq
If nothing is invested in derivative ($\eta=0$) the
fraction invested in stock ($\rho=\frac{\mu}{\sigma^2}$) is the same
as the optimal investment fraction in the stock-and-bond problem.
When \eq{BS:blackscholes} holds true any dynamics along 
the {\em marginal subspace} does not produce any gain or loss.
That means we can invest $\rho\,W$ (in value) in stock and 
$-\frac{\sigma}{\sigma_{d}}\rho\,W$ (in value) in derivative, 
for any $\rho$. Expressed in stock price $\psi_{t}$  
and numbers of stock, $n_{\psi}$, the value invested in stock is
$n_{\psi} \psi_{t}$. The value invested in derivative
is hence $-\frac{\sigma}{\sigma_d} n_{\psi} \psi_{t}$,
but also $n_d C$, if $n_d$ is the
numbers of derivative. The number of stock per derivative is hence 
\beq
\frac{n_S}{n_d} = - \frac{\sigma_d}{\sigma} \frac{C}{\psi} 
= - \de_{\psi} C 
\eeq
The portfolio along the marginal subspace is hence a simply Black-Scholes 
delta hedge
\beq
\Delta:=\de_{\psi} C 
\eeq
following the standard financial notation.

\section{Hamilton-Bellman-Jacobi problem for analytic transaction costs}
\label{sec:HJB}

The use of analytic transaction costs renders the Hamilton-Jacobi-Bellman 
problem simpler to study. In the frame of reference fixed by the eigenvectors 
of the Hessian of the utility function, the stochastic dynamics is governed 
by the system of stochastic differential equations
\beq
&&dW= W\sbr{\mu \zeta\, dt+\sigma \zeta\,dB_{t}}-
W\mathcal{F}\rbr{f_{\zeta},f_{\vartheta}}dt
\label{HJB:wealth}
\\
&&d\zeta =\sbr{f_{\zeta}+a_{\zeta}+
\zeta\,\mathcal{F}\rbr{f_{\zeta},f_{\vartheta}}}dt
+b_{\zeta}\, dB_{t}
\label{HJB:relevant}
\\
&&d\vartheta =
\sbr{f_{\vartheta}+a_{\vartheta}+\vartheta\,
\mathcal{F}\rbr{f_{\zeta},f_{\vartheta}}}\,dt
+b_{\vartheta}\, dB_{t}
\label{HJB:marginal}
\\
&&d\sigmad=\sigma\, H\,dt+\sigma\,K\,dB_{t}
\label{HJB:sigmad}
\\
&&d\psi =\mu \psi\,dt+\sigma\,\psi\, dB_{t}
\label{HJB:price}
\eeq
The drift and diffusion fields in these coordinates are
\beq
&&a_{\zeta}=(\mu-\sigma^2\zeta)
\frac{\sigma^3\,\zeta (1-\zeta)+\sigmad^2\zeta(\sigmad-\sigma\zeta)+
\sigma \sigmad\vartheta(\sigma-\sigmad) }
{\sigma (\sigma^{2}+\sigmad^{2})}
\nn
&&\qquad+
\sigma\frac{\sigmad\zeta-\sigma\vartheta}{\sigma^{2}+\sigmad^{2}}
\sbr{H+K\rbr{\sigmad-\sigma\zeta}}
\label{HJB:portfoliodrift}
\\
&&b_{\zeta}=\frac{\sigma^3\,\zeta\,(1-\zeta)+\sigmad^2\zeta(\sigmad-\sigma\zeta)+
\sigma \sigmad\vartheta(\sigma-\sigmad) }
{(\sigma^{2}+\sigmad^{2})}+
\sigma\,\frac{\sigmad\zeta-\sigma\vartheta}{\sigma^{2}+\sigmad^{2}}\,K
\label{HJB:portfoliodiffusion}
\\
&&a_{\vartheta}=(\mu-\sigma^2\zeta)
\sbr{\frac{\sigmad\zeta(\sigma-\sigmad)}{\sigma^{2}+\sigmad^{2}}+
\vartheta\rbr{\sigmad\,\frac{\sigma+\sigmad}{\sigma^{2}+\sigmad^{2}}-\zeta}}
\nn
&&\qquad+
\sigma\frac{\sigma\zeta+\sigmad\vartheta}{\sigma^{2}+\sigmad^{2}}
\sbr{H+\sigma\rbr{1-\zeta}K}
\label{HJB:marginaldrift}
\\
&&b_{\vartheta}=\sigma \sbr{\frac{\sigmad\zeta(\sigma-\sigmad)}
{\sigma^{2}+\sigmad^{2}}+
\vartheta\rbr{\sigmad\,\frac{\sigma+\sigmad}{\sigma^{2}+\sigmad^{2}}-\zeta}}
+\sigma \frac{\sigma\zeta+\sigmad\vartheta}{\sigma^{2}+\sigmad^{2}}K
\label{HJB:marginaldiffusion}
\eeq
while the time change of $\sigma_d$ is expressed in terms of
two new amplitudes:
\beq
&&H=\frac{1}{\sigma}\sbr{\de_{t}\sigma_{d}+\mu\,\psi\,\de_{\psi}\sigma_{d}+
\frac{\sigma^2\,\psi^2}{2}\de_{\psi}^2\sigma_{d}}
\nn
&&K=\psi\,\de_{\psi}\sigma_{d}
\label{HJB:sigmadfields}
\eeq
By using the Black and Scholes equation, $H$ can be expressed in
terms of $K$ and the other parameters:
\beq
H=\sigma K \rbr{\frac{\mu}{\sigma^2}+\frac{\sigmad}{\sigma}}
\eeq
The dynamic programming equation for quadratic friction 
\beq
\mathcal{F}(f_{\zeta}\,,f_{\vartheta})=
\gamma\frac{\sigma^{4}}{\mo{h_{s}}^{2}}
\rbr{f_\zeta+\frac{\sigmad}{\sigma}f_\vartheta}^{2}+
\gamma_{d}\frac{\sigma^{4}}{\mo{h_s}^{2}}
\rbr{\frac{\sigmad}{\sigma} f_{\zeta}- f_{\vartheta}}^{2}
\eeq
is
\beq
&&\de_t\lambda+  \sbr{f_{\zeta}+a_{\zeta}+
z\,\mathcal{F}(f_{\zeta}\,,f_{\vartheta})}\de_z \lambda+ 
\sbr{f_{\vartheta}+a_{\vartheta}+
y\,\mathcal{F}(f_{\zeta}\,,f_{\vartheta})} \de_y \lambda
\nn
&&+\mu p\de_{p}\lambda+\frac{b^2_{\zeta}}{2}\de_z^2 \lambda+
\frac{b^2_{\vartheta}}{2}\de_y^2 \lambda+\frac{\sigma^2 p^2}{2}\de_{p}^2\lambda+
b_{\zeta} b_{\vartheta}\de_{z\,y}\lambda+b_{\zeta}\,\sigma p\de_{z p}\lambda+
b_{\vartheta}\sigma p\de_{y p}\lambda
\nn
&&+\mu z-\mathcal{F}(f_{\zeta},f_{\vartheta})=0
\label{HJB:dp}
\eeq
The optimal investment strategy is sought by requiring the capital
growth $\lambda$ reach as a functional of the controls 
$(f_{\zeta}\,,f_{\vartheta})$ a stationary point:
\begin{eqnarray}
&&-\pder{\mathcal{F}}{f_{\zeta}}
+ \rbr{1 + z \pder{\mathcal{F}}{f_{\zeta}}} 
\de_z \lambda+ \pder{\mathcal{F}}{f_{\vartheta}}\,y \de_y\lambda =0 
\nonumber\\
&&-\pder{\mathcal{F}}{f_{\zeta}} + z \pder{\mathcal{F}}{f_{\vartheta}}
\de_z\lambda + \rbr{1 + \,\pder{\mathcal{F}}{f_{\vartheta}}\, y} \de_y \lambda=0 
\end{eqnarray}
The stationary point equations admit a unique solution for the 
stochastic controls
\beq
\label{HJB:stoch-control}
&&f_{\zeta}=\frac{\rbr{\gamma \sigmad^2+\gammad\,\sigma^2}\de_{z}\lambda+
\sigma\sigmad (\gammad-\gamma)\de_{y}\lambda}
{2\,\gamma\,\gammad\,\sigma^2\,\rbr{1-z\de_{z}\lambda-y\de_{y}\lambda}}
\nn
\label{HJB:stoch-control-d}
&&f_{\vartheta}=\frac{\rbr{\gamma \sigma^2+\gammad\,\sigmad^2}\de_{y}\lambda+
\sigma\sigmad (\gammad-\gamma)\de_{z}\lambda}
{2\,\gamma\,\gammad\,\sigma^2\,\rbr{1-z\de_{z}\lambda-y\de_{y}\lambda}}
\eeq
which inserted in the dynamic programming equation yield the
the Hamilton-Jacobi-Bellman equation for the model
\beq
&&\de_t\lambda+ a_{\zeta}\de_z \lambda+ a_{\vartheta}\de_y \lambda
+\mu p\de_{p}\lambda+\frac{\gammad\,(\sigma \de_z \lambda+
\sigmad \de_y \lambda)^2+\gamma\,(\sigmad \de_z \lambda-\sigma \de_y \lambda)^2}
{4\, \sigma^2\gamma\gammad\rbr{1-z\de_{z}\lambda-y\de_{y}\lambda}}
\nonumber\\
&&\quad+\frac{b_{\zeta}^2}{2}\de_z^2 \lambda+\frac{b_{\vartheta}^2}{2}\de_y^2 \lambda
+\frac{(\sigma p)^2}{2}\de_p^2 \lambda
+b_{\zeta} b_{\vartheta}\de_{z\,y}\lambda+b_{\zeta}\sigma p\de_{z p}\lambda+
b_{\vartheta}\sigma p\de_{y p}\lambda
\nonumber\\
&&\quad+\frac{\mu^2}{2\sigma^2}- \frac{\sigma^2}{2}\rbr{\frac{\mu}{\sigma^2}-z}^2=0
\label{HJB:hjb}
\eeq
The dynamics is fully specified by associating to \eq{HJB:hjb} the equation for
the derivative volatility
\beq
\de_{t} \sigmad+\frac{\sigma\,p\,}{2}\de_{p} \sigmad^2+
\frac{\sigma^2 p^2}{2}\de^2_{p}\sigmad=0
\eeq
Equation~\eq{HJB:hjb} contains in principle the solution to our problem.
The expected capital growth rate $\lambda$ determines the stochastic controls
through (\ref{HJB:stoch-control}).
Qualitative and quantitative analyses of the control strategies are pursued
in sections~\ref{sec:qualitative} and ~\ref{sec:quantitative} below.

The Hamilton-Jacobi-Bellman equation should be supplemented by boundary 
conditions on $\lambda$.
Arguably, the most natural would be to additionally assume that the
process is confined to some domain, and hence that the normal gradient
of $\lambda$ vanishes on the boundaries of that domain. For the rest of this paper,
we will be concerned with a description close to the optimum. We will hence
assume that the boundaries are far away, that the probability is there
small, and that we will not need to further specify the boundary conditions.

\section{Qualitative analysis of the Hamilton-Jacobi-Bellman equation}
\label{sec:qualitative}

Qualitatively, one can imagine the following scenario.
As the utility changes with the position along the stable
direction, one may postulate a {\it fast control}
along that direction, striving to bring the position
close to the marginal. Depending on where one is along
the marginal, there would then be different (expected)
transaction costs in keeping the position in the stable
direction close to zero.
Hence, all positions along the marginal are not equivalent,
because they lead to different (expected) transaction costs
in the stable direction. In fact, we can then posit that the preferred
position along the marginal is such that
$b_{\zeta}$ of \eq{HJB:portfoliodiffusion}, evaluated at
\beq
\zeta^{\star}=\frac{\mu}{\sigma^2}
\label{qualitative:optimum}
\eeq
is as small as possible. Since this function is linear in the
marginal coordinate (see below), we then have a 
prescription for the optimum allocation into a Black-Scholes
hedge, as function of time and price of the underlying
\beq
b_{\zeta}^{\star}=D+\rbr{\sigmad\zeta^{\star}-\sigma \vartheta^{\star}}
\frac{\rbr{\sigmad-\sigma}\sigmad+\sigma\,K}{\sigma^{2}+\sigmad^{2}}
\label{qualitative:diffusion}
\eeq
with  
\beq
D:=\frac{\mu}{\sigma}\,\rbr{1-\frac{\mu}{\sigma^2}}
\eeq
the diffusion amplitude in the absence of derivative trading. From the
definition of $K$ (see appendix~\ref{ap:greeks}) the right hand side of
\eq{qualitative:diffusion} can also be rewritten as
\beq
b_{\zeta}^{\star}=D-\frac{\sigmad\zeta^{\star}-\sigma \vartheta^{\star}}
{\sigma^{2}+\sigmad^{2}}\frac{2\,\de_{t}C}{C}
\label{qualitative:diffusion2}
\eeq
Note that the time variation of the derivative price is usually denoted
in the financial literature by the Greek letter $\Theta$. The relations
satisfied by the derivative-$\Theta$ with other commonly used financial
indicators as the derivative $\Delta$ and $\Gamma$ are recalled in 
appendix~\ref{ap:greeks}.
The condition
\beq
b_{\zeta}^{\star}=0
\label{qualitative:nodiffusion}
\eeq
is enforced by setting
\beq
\vartheta^{\star}=\frac{\sigmad}{\sigma}\zeta^{\star}-
\frac{\rbr{\sigmad^2+\sigma^2}\,D\,C}{2\,\sigma\,\de_{t}C}
\label{qualitative:thetastar}
\eeq 
This equation is in one sense the main result of the present
work. It is therefore useful to rewrite it directly in the
original variables, i.e. the fractions invested in stock and derivative:
\beq
&&\rho^{\star} = 
\frac{\sigma}{\sigma^2 + \sigma_d^2}
(\sigma\zeta^{\star}+\sigma_d \vartheta^{\star})
= \frac{\mu}{\sigma^2} - 
\frac{\sigmad\,D\,C}{2\,\de_{t}C}= \frac{\mu}{\sigma^2} + 
\frac{\Delta\,D}{\sigma\,p\,\Gamma}
\nn
&&\eta^{\star} = 
\frac{\sigma}{\sigma^2 + \sigma_d^2}
(\sigma_d\zeta^{\star}-\sigma \vartheta^{\star}) = 
\frac{D\,\sigma\,C}{2\,\de_{t}C}
= - \frac{\Delta\,D}{\sigmad\,p\,\Gamma}
\label{qualitative:main}
\eeq 
A consequence of these equations is that if
$\sigma_d$ diverges, $\rho^{\star}$ stays finite, while
$\eta^{\star}$ tends to zero ($\theta^{\star}$ diverges as $\sigma_d$).
This happens for European Call Options close to expiry and 
at-the-money (see appendix, $\sigma_d$ is then proportional to 
the ``Greek'' $\Delta$, and inversely proportional to the option price C).
Following Black-Scholes hedging directly can then lead to
large transaction costs, because of ``portfolio flapping''
(switching between the fully hedged and totally unhedged
positions in response to small changes in the underlying).
We see that from the perspective of optimal investment
strategies, this problem does not appear, since 
such portfolios do not contain any at-the-money options close to expiry.  
 
In the opposite limit of a large investment horizon, the derivative 
volatility tends to the volatility of the underlying. Furthermore the
inequality
\beq
\left|\frac{\de_{t}\,C}{C}\right|\,\ll\,1
\eeq
holds true requiring increasingly large investments in derivatives 
in order to enforce \eq{qualitative:nodiffusion}. In such limit 
the optimal size of the investment in stocks is also seen to diverge.  
The reason is that the drift \eq{HJB:portfoliodrift} and diffusion amplitude 
\eq{HJB:portfoliodiffusion} along the stable direction become independent 
of the marginal direction if the infinite horizon limit is taken for any 
arbitrary finite value of $\theta$. In particular \eq{HJB:portfoliodrift},
\eq{HJB:portfoliodiffusion} become in the portfolio variable $\zeta$ identical 
to the drift and diffusion amplitude felt in the stock and bond model studied
in \cite{AurellMG2}. These phenomena indicate a break-down of the argument 
used to derive \eq{qualitative:nodiffusion}.
Qualitatively one expects in this second limit the optimal investment strategy
to treat the stock and the derivative in a similar manner analogous to the one
depicted in \cite{AurellMG2}.

\section{Systematic analysis of the Hamilton-Jacobi-Bellman equation}
\label{sec:quantitative}

In this section we will use a systematic multi-scale
analysis to the Hamilton-Jacobi-Bellman equation. We will show that 
the previous qualitative analysis is well founded.
Furthermore, we are also able to treat the case when
the (putative) optimal position $\theta^{\star}$
diverges, and, more generally, we can compute the
control to be exerted on $\zeta$ and $\theta$.
Nevertheless, the main interest here is conceptual,
that the previous analysis can be systematically justified.

To start with it is convenient to write the logarithmic growth
in the form
\beq
\lambda=\frac{\mu^2}{2\,\sigma^2}(T-t)+\varphi
\eeq
The first term corresponds to growth in the absence of transaction 
costs. The intensity of transaction costs is then measured by the
two adimensional parameters
\beq
\eps=\sigma^2\frac{\gamma\,\gammad}{\gamma+\gammad}
\eeq
and
\beq
\flu{\gamma}=\frac{\gammad-\gamma}{\gamma+\gammad}
\eeq
In order to construct an asymptotic expansion around the ideal case of 
no transaction costs it is convenient to shift the origin of the coordinates
$(z,y)$ to 
\beq
z \Rightarrow \zeta^{\star}+z\,,
\qquad
y \Rightarrow \vartheta^{\star}+y
\eeq
The Hamilton-Jacobi-Bellman equation can be rewritten as
\beq
&&\de_t\varphi+ a_{\zeta}\de_z \varphi+ a_{\vartheta}\de_y \varphi+
\mu p\de_{p}\varphi+ 
\frac{ (\sigma^2+\sigmad^2)}{8\eps }\frac{(\de_z \varphi)^2+(\de_y \varphi)^2}
{1-(\zeta^{\star}+z)\de_{z}\varphi-(\vartheta^{\star}+y)\de_{y}\varphi}
\nn
&&+\frac{\flu{\gamma}}{8 \eps}
\frac{(\sigma \de_z \varphi+\sigmad \de_y \varphi)^2-
(\sigmad \de_z \varphi-\sigma \de_y \varphi)^2}
{1-(\zeta^{\star}+z)\de_{z}\varphi-y\de_{y}\varphi}+
\frac{b_{\zeta}^2}{2}\de_z^2 \varphi+\frac{b_{\vartheta}^2}{2}\de_y^2 \varphi
+\frac{(\sigma p)^2}{2}\de_p^2 \varphi
\nn
&&+b_{\zeta} b_{\vartheta}\de_{z\,y}\varphi+b_{\zeta}\sigma p\de_{z p}\varphi+
b_{\vartheta}\sigma p\de_{y p}\varphi- \frac{\sigma^2\,z^2}{2}=0
\label{quantitative:hjb}
\eeq
By dimensional analysis, one can motivate the following choice
of scales in time, stable subspace ($z$) and marginal subspace ($y$):
\bea
\varphi=\varepsilon\,\sum_{n=0}^{\infty} \varepsilon^{\frac{n}{8}}
\phi_{(n)}\rbr{\frac{t}{\varepsilon^{1/2}},
\frac{t}{\varepsilon^{3/8}},\frac{t}{\varepsilon^{1/4}},
\frac{t}{\varepsilon^{1/8}},\dots,\frac{z}{\varepsilon^{1/4}},
\frac{z}{\varepsilon^{1/8}},\dots,\frac{y}{\varepsilon^{1/8}},\dots,t,p}
\label{quantitative:ansatz}
\eea
Introducing fast and slow variables
\bea
\begin{array}{lll}
\ta:=\frac{t}{\varepsilon^{1/2}}\,, \quad &\quad 
\tb:=\frac{t}{\varepsilon^{3/8}}\,, 
\quad & \quad  \dots 
\\
\za:=\frac{z}{\varepsilon^{1/4}}\,,  \quad & \quad 
\zb:=\frac{z}{\varepsilon^{1/4}}\,, 
\quad & \quad \dots
\\
\za:=\frac{y}{\varepsilon^{1/8}}\,,  \quad & \quad \dots
\end{array}
\eea
derivatives are rewritten as
\beq
&&\de_{t}\varphi=\eps^{1/2}\,\de_{\ta}\phi_{(0)}+\eps^{5/8}\,\sbr{\de_{\tb}\phi_{(0)}
+\de_{\ta}\phi_{(1)}}
\nn
&&\qquad+\eps^{3/4}\,\sbr{\de_{\tc}\phi_{(0)}+\de_{\tb}\phi_{(1)}
+\de_{\ta}\phi_{(0)}}+\dots
\nn
&&
\de_{z}\varphi=\eps^{3/4}\,\de_{\za}\phi_{(0)}+\eps^{7/8}\,
\sbr{\de_{\zb}\phi_{(0)}+\de_{\za}\phi_{(1)}}+\dots
\nn
&&
\de_{y}\varphi=\eps^{7/8}\,\de_{\ya}\phi_{(0)}+\dots
\eeq
We introduce ``slower'' variables in the perturbative expansion for the 
following reason. A general feature of the expansion we are considering 
is that lower order approximate solutions enter the linear
partial differential equations governing higher order ones in the form of 
non-homogeneous terms. By Fredholm's alternative it follows that the 
perturbative expansion is consistent if and only if these non-homogeneous term 
have no overlap with kernel of the linear differential operator associated to 
the homogeneous part of the equations. According to the standard multiscale 
method (see for example \cite{Bocquet}) slower variables can be used
to enforce the consistency conditions. 
The hierarchy of perturbative equations starts with 
\beq
&\eps^{1/2}:&\quad \de_{\ta}\phi_{(0)}+
\frac{(1+\flu{\gamma})\sigma^2+(1-\flu{\gamma})\sigmad^2}{8}\,
\sbr{\de_{\za}\phi_{(0)}}^2
+\frac{{b^{\star}_{\zeta}}^2}{2}\de_{\za}^2\phi_{(0)}
\nn
&&\qquad\quad
-\frac{\sigma^2 \za^2}{2}=0
\label{quantitative:hierarchy1}
\\
&\eps^{5/8}:&\quad L(\ta,\za,\phi_{0})\phi_{(1)}
+\frac{b_{\zeta}^{\star}}{2}\de^2_{\za^2}\phi_{(1)}=
-L(\tb,\zb,\phi_{0})\phi_{(0)}
\nn
&&
\qquad\quad
-\frac{\sigma\,\sigma_d}{2}\rbr{\de_{\za}\phi_{(0)}}\rbr{\de_{\ya}\phi_{(0)}}
-b^{\star}_{\zeta}\,b^{\star}_{\vartheta}\,\de_{\za\ya}^2\phi_{(0)}
\nn
&&
\qquad\quad
-b_{\zeta}^{\star}\,\sigma\,\de^2_{\za\zb}\phi_{(0)}
\label{quantitative:hierarchy2}
\\
&\eps^{3/4}:&\quad \dots
\nonumber
\eeq
where $b^{\star}_{\zeta}$, $b^{\star}_{\vartheta}$ are evaluated at $z$ equal
$\zeta^{\star}$ and depend parametrically upon $\vartheta^{\star}$ whilst
\beq
L(t,z,\phi_{0}):=\pder{}{t}+\frac{(1+\flu{\gamma})\,\sigma^2\,+\,
(1-\flu{\gamma})\,\sigmad^2}{2}\rbr{\pder{\phi_{(0)}}{\za}} 
\,\pder{}{z}
\eeq

\section{Leading order asymptotics}
\label{sec:leading}

Formally the leading order of the perturbative hierarchy of equations
coincide with the one of the stock and bond market model studied in
\cite{AurellMG1}. Setting
\beq
A=\frac{(1+\flu{\gamma})\,\sigma^2+(1-\flu{\gamma})\,\sigmad^2}{2}
\eeq
for $z$ sufficiently small the asymptotic expression of the logarithmic 
growth of the investor capital is \cite{AurellMG2}
\beq
\lambda(z,t;T)&=&\rbr{\frac{\mu^2}{2\,\sigma^2}
-\eps^{1/2}\frac{{b^{\star}_{\zeta}}^2}{2}}\,(T-t)-\eps^{1/2}\,
\rbr{\frac{\sigma^2}{A}}^{1/2}\,\frac{z^2}{2}
\nn
&+&
\frac{2\,{b^{\star}_{\zeta}}^2\,\eps}{A}
\ln\cbr{\sqrt{2}\sum_{n=0}^{\infty}\,\frac{e^{-2\,n\,\sqrt{\frac{\sigma^2\,A}{2}}
\frac{T-t}{\eps^{1/2}}}}{2^{2\,n}\,\Gamma(n+1)}H_{2n}
\rbr{\rbr{\frac{\sigma^2\,A}{2\,{b^{\star}_{\zeta}}^4}}^{1/4}
\frac{z}{\eps^{1/4}}}}
\nn
&+&O(\eps^{1+1/8}\phi_{(1)})
\eeq
with $H_{2\,n}$ denoting the Hermite polynomial of order $2\,n$. The argument
of the logarithm can be further resummed using the Fourier representation of the 
generating function of the Hermite polynomials
\beq
H_{n}(a x)=(-1)^n\,e^{a^2 x^2}\eval{\der{^{n}e^{y^2}}{y^{n}}}{y=a x}=e^{a^2 x^2}
\int \frac{dp}{\sqrt{2 \pi}}\,(-\imath p)^n\,e^{\imath p a x-\frac{p^2}{4}}
\eeq
The result is
\beq
\lambda(z,t;T)&=&\rbr{\frac{\mu^2}{2\,\sigma^2}
-\eps^{1/2}\frac{{b^{\star}_{\zeta}}^2}{2}}\,(T-t)
-\eps^{1/2}\,
\rbr{\frac{\sigma^2}{A}}^{1/2}\,\frac{z^2}{2}\tanh
\cbr{\sqrt{\frac{\sigma^2\,A}{2}}\frac{T-t}{\eps^{1/2}}}
\nn
&+&\frac{{b^{\star}_{\zeta}}^2\,\eps}{A}
\ln\cbr{\frac{2}{1+e^{-2\,\sqrt{\frac{\sigma^2\,A}{2}}\frac{T-t}{\eps^{1/2}}}}}
\nn
&+&O(\eps^{1+1/8}\phi_{(1)})
\label{leading:result}
\eeq
At variance with \cite{AurellMG2} the diffusion coefficient ${b^{\star}_{\zeta}}^2$ 
in \eq{leading:result} depends for any finite investment horizon upon 
$\theta^{\star}$. The logarithmic growth $\lambda$ attains 
a maximum for $z$ equal to zero corresponding to the optimal portfolio 
in the absence of transaction costs. 
The value of this maximum defines the growth rate of the investor capital. 
It is straightforward to verify that the conditions 
\eq{qualitative:nodiffusion}, \eq{qualitative:thetastar} specify 
the {\em supremum} for the growth rate of the investor capital.
The overall logarithmic growth becomes in such a limit 
\beq
\lambda(z,t;T)&\overset{b^{\star}_{\zeta}\to 0}{\to}&
\frac{\mu^2}{2\,\sigma^2}\,(T-t)
-\eps^{1/2}\,
\rbr{\frac{\sigma^2}{A}}^{1/2}\,\frac{z^2}{2}\tanh
\cbr{\sqrt{\frac{\sigma^2\,A}{2}}\frac{T-t}{\eps^{1/2}}}
\nn
&+&O(\eps^{1+1/8}\phi_{(1)})
\label{leading:result2}
\eeq
The qualitative conclusion that can be inferred from \eq{leading:result2}
is that the inclusion in the optimal portfolio of a derivative product  
quells the effect of transaction costs from the capital growth rate.

The mathematical conditions for the validity of the asymptotic expression 
\eq{leading:result2} of the logarithmic growth are determined by 
\eq{qualitative:main}. The corresponding portfolio is well defined close 
to maturity and for values of the underlying price close to the strike price, 
when the volatility of the derivative price becomes very large.
It is also worth stressing that the asymptotics \eq{leading:result2} holds true 
for values of $z$ sufficiently small that the effect of the boundary conditions 
can be neglected:
\beq
z \ll 1
\eeq
The reasoning allowing to derive the asymptotics \eq{leading:result2} from
\eq{leading:result} breaks down in the large investment horizon limit as 
discussed at the end of section~\ref{sec:qualitative}. Namely in such a limit 
the terms proportional to $\vartheta^{\star}$ in \eq{quantitative:hierarchy1}
vanish, leaving with an equation in the portfolio variable $z$ of the same form
of the one describing the investment strategy in the absence of derivatives
the solution whereof was studied in \cite{AurellMG2}. 

The analysis of the intermediate dynamical regime between maturity and large
horizon requires to take into account the boundary conditions associated to the 
Hamilton-Jacobi-Bellman equation \eq{HJB:hjb} and is beyond the scope of this 
paper.

\subsection{Corrections to the leading order}

Inspection of \eq{quantitative:hierarchy2} shows that it is consistent to set
\beq
\phi_{(1)}=0
\eeq
with $\phi_{(0)}$ independent of the first set of slower variables. Hence
the first non-trivial correction to \eq{leading:result2} turns out to be of 
the order $O(\eps^{1+2/8}\phi_{(2)})$ as in the case of a market model
without derivative products \cite{AurellMG2}. 

\section{Conclusions}
We have shown that optimum investment strategies in a portfolio
of stocks, bond and a derivative can be determined by Hamilton-Jacobi-Bellman
techniques. Black-Scholes equation appears as a solvability condition
for the problem to be well-founded. Optimal strategies can be described
as ``fuzzy Black-Scholes'': if transaction costs are small, optimal
portfolios are not far from Black-Scholes delta hedges.
\\\\
We believe it of conceptual interest that Black-Scholes pricing emerges
as a solvability condition for an ensemble of possible investment strategies.
Hence, Black-Scholes has been motivated in a weaker setting, where there
is no replicating portfolio.
Second, we have shown that expected transaction costs can be lowered
by choosing between investments in both stocks and derivatives, and not
only in stock. This is not surprising, but the point has not previously
been made previously by systematic analysis, to our best knowledge. We note that
the qualitative analysis can be extended to the case of several derivatives
on the same stock. Although there is a ``law of diminishing returns'', expected
transaction costs can then be lowered further.

Finally we have made explicit the optimal fraction invested in derivatives
in terms of the standard financial ``Greeks''.

\appendix

\section{European call option}
\label{ap:option}
\setcounter{equation}{0}
\renewcommand{\theequation}{A.\arabic{equation}}

The boundary condition associated to Black and Scholes's equation
for an European call option is
\beq
C(p,0)=\mathrm{max}\cbr{p-\bar{p},0}
\eeq
with $\bar{p}$ the exercise (strike) price. The solution at zero discount rate 
is
\beq
C(p,T-t)=p N(\phi_{1})-\bar{p} N(\phi_{2})
\eeq
where
\beq
N(x)=\int_{-\infty}^{x}\!\!\!dy\,\frac{e^{-\frac{y^2}{2}}}{\sqrt{2\,\pi}}=
\frac{1}{2}\sbr{1+\mathrm{Erf}\rbr{\frac{x}{\sqrt{2}}}}
\eeq
and
\beq
\phi_{1}:=\frac{\ln\rbr{\frac{p}{\bar{p}}}+\frac{\sigma^2\,(T-t)}{2}}
{\sigma\,\sqrt{T-t} }
\,,\qquad\qquad
\phi_{2}:=\phi_{1}-\sigma\,\sqrt{T-t}
\eeq
Observing that
\beq
\pder{\phi_{1}}{p}=\pder{\phi_{2}}{p}
\eeq
it is found that for an European call option
\beq
\sigma_{d}=\frac{\sigma\,p\,N(\phi_1)}{p\,N(\phi_1)-\bar{p}\,N(\phi_2)}
\eeq
Thus, in the large investment horizon limit $T-t\uparrow \infty$ and 
in the limit of underlying prices much larger than the strike at maturity date
$p\,\ll\,\bar{p}$ the volatility of the derivative product tends to the 
volatility of felt by the underlying.

\section{Relation with the ``Greeks''}
\label{ap:greeks}
\setcounter{equation}{0}
\renewcommand{\theequation}{B.\arabic{equation}}

The sensitivity of the derivative price to variation of the underlying
are measured by the ``Greeks'': a set of factor sensitivities used 
extensively by traders to quantify the exposures of portfolios that contain 
options. In the present case the Greeks of relevance are
\beq
\Delta:=\pder{C}{p}\,,
\qquad
\Gamma:=\pder{^2C}{p^2}\,,
\qquad
\Theta:=\pder{C}{t}
\eeq
In such a case
\beq
\sigma_{d}=\sigma\frac{p\Delta}{C}
\eeq
From the definition \eq{HJB:sigmadfields} of the field $K$ 
and the Black and Scholes equation \eq{BS:BS} it follows 
\beq
K&=&\sigma \frac{p \Delta}{C}-\sigma \frac{p^2\Delta^2}{C^2}
+\sigma\frac{p^2\Gamma}{C}=
\sigmad\,\rbr{1-\frac{\sigmad}{\sigma}}+\sigmad\frac{ p \Gamma}{\Delta}
\nn
&=&\sigmad\,\rbr{1-\frac{\sigmad}{\sigma}}-\frac{ 2  \Theta}{\sigma C}
\label{greeks:kappa}
\eeq
For an European call options 
\beq
 \Theta=-\sqrt{\frac{p\,\bar{p}}{2\,\pi}}
\frac{e^{-\frac{(\ln\frac{p}{\bar{p}})^2}{2\,\sigma^2\,(T-t)}-
\frac{\sigma^2}{8}(T-t)}}{2\,\sqrt{T-t}}
\eeq
the fields $K$ tends to a distribution when $t$ tends to the maturity
date.

\section{Asymptotics of the probability distribution of the stock investment}
\label{ap:probability}
\setcounter{equation}{0}
\renewcommand{\theequation}{C.\arabic{equation}}

The leading order asymptotics to the Hamilton Bellman Jacobi equation
can be written as
\beq
\de_{t}\chi+\frac{A}{4\,\eps}\,\rbr{\de_{z}\chi}^2
+\frac{\nu}{2}\de_{z}^2\chi
+\frac{\mu^2}{2\,\sigma^2}-\frac{\sigma^2 z^2}{2}=0
\label{probability:potential}
\eeq
It describes  the evolution equation of the potential of a drift field
\beq
v=\frac{A}{2\,\eps}\de_{z}\chi
\eeq
advecting the Fokker-Planck equation
\beq
\de_{t}P+\de_{z}(v\,P)-\frac{\nu}{2}\de^2\,P=0
\eeq
describing within approximation the probability density of the investment
in stock.
The general solution can be written in path integral form
\beq
P(z^{\prime},T|z,t)=\int 
\mathcal{D}[\zeta_{s}]\,\delta(\zeta_{t}-z)\delta(\zeta_{T}-z^{\prime})\,e^{-\ac}
\eeq
with
\beq
\ac=\frac{1}{2\,\nu}\int_{t}^{T}ds\,\cbr{\sbr{\dot{\zeta}-
\frac{A}{2\,\eps} \de_{\zeta}\chi}^2
+\frac{\nu A}{2\,\eps} \de_{\zeta}^2\chi}
\eeq
The path integral can be performed exactly (see \cite{MG1} for details) 
\beq
P(z^{\prime},T|z,t)=e^{\frac{A\,\mu^2\,(T-t)}{2\,\nu\,\sigma^2\,\eps}-
\frac{A\,\chi(z,t)}{2\nu\,\eps}}\,
\frac{e^{-\omega\frac{(z^2+z^{\prime\,2})\cosh\cbr{\omega(T-t)}-2 z\,z^{\prime}}
{2\,\nu\,\sinh\cbr{\omega(T-t)}}}}{\sqrt{2\,\pi\,\nu\,\sinh \omega (T-t)}}
\label{probability:probability}
\eeq
having used the boundary condition
\beq
\chi(z,T)=0
\eeq
and the notation
\beq
\omega=\sqrt{\frac{A\sigma^2}{2\,\eps}}
\eeq
The explicit form of $\chi$ is obtained by imposing probability 
conservation over $z^{\prime}$. If this latter variable takes values
on the entire real axis, the result \eq{leading:result} given in the main text 
as leading asymptotic to the full solution is recovered. 
The corresponding form of the probability distribution is
\beq
P(z^{\prime},T|z,t)=\sqrt{\frac{\omega}{2\,\pi\,\nu\,\tanh \omega (T-t)}}
\,e^{-\omega\frac{(z^{\prime}-z/\cosh\cbr{\omega(T-t)})^2}
{2\,\nu\,\tanh \omega (T-t)}}
\label{probability:asympt}
\eeq 
A direct calculation allows to verify that 
\eq{probability:asympt} satisfies the equality
\beq
\chi(z,t)=\int_{t}^{T}ds\int_{\mathbb{R}} dy \sbr{\frac{\mu^{2}}{2\,\sigma^2}
-\frac{\sigma^2\,z^2}{2}} P(y,s|z,t)
\eeq
with $\chi(z,t)$ also given by \eq{leading:result} as required by the
stochastic dynamics underlying the Hamilton-Jacobi-Bellman equation.

\section*{Acknowledgments}
\noindent
The authors are pleased to acknowledge discussions with A. Kupiainen and
A. Vulpiani during the preparation of this work.
This work was supported by the Swedish Research Council 
(E.A.), and by the Centre of Excellence 
{\em Geometric Analysis and Mathematical Physics} by the Department of 
Mathematics and Statistics of the University of Helsinki (P.M.G.).


\begin{thebibliography}{00}




\bibitem{AtkinsonPliskaWilmott}
Atkinson, Pliska and P. Wilmott 
``Portfolio management with transaction costs'', Proc. R. Soc. Lond. {\bf A 453} 
551-562 (1997).

\bibitem{AvellanedaParas} 
M. Avellaneda and A.~Paras
``Dynamic Hedging Portfolios for Derivative Securities in the 
Presence of Large Transaction Costs'',
Appl. Math. Finance  {\bf 1}, 165-193 (1994).

\bibitem{AurellMG1} E.~Aurell and P.~Muratore-Ginanneschi,
``Financial Friction and Multiplicative Markov Market Game'', 
{\it International J. of Theoretical and Applied 
Finance} (IJTAF) {\bf 3}, 501-510 (2000)  and cond-mat/9908253.

\bibitem{AurellMG2} E.~Aurell and P.~Muratore-Ginanneschi,
``Growth-optimal strategies with quadratic friction over
finite-time investment horizons'', 
{\it International J. of Theoretical and Applied 
Finance} (IJTAF) {\bf 7},  645-657 (2004) and cond-mat/0211044.



\bibitem{BensaidLesnePagesScheinkman}
B.~Bensaid, J.-P.~Lesne, H.~Pages and J.~Scheinkman, 
``Derivative Asset Pricing With Transaction Costs'',
Banque de France - Direction Generale des Etudes
Papers {\bf 15}, (1991).

\bibitem{Bjork}
T.~Bj{\"o}rk
{\it Arbitrage Theory in Continuous Time}, Oxford University Press 
2nd ed. (2004).

\bibitem{BoyleVorst} P.P.~Boyle and T.~Vorst
 ``Option Replication in Discrete Time with Transaction Costs'',
{\bf 47}, 271-293 (1992).

\bibitem{Bocquet}  L.~Bocquet,
``High friction limit of the Kramers equation: the multiple
time scale approach'', American Journal of Physics {\bf 65}
(1997), 140-144 and cond-mat/9605186.

\bibitem{BouchaudSornette}
J.-P.~Bouchaud and D.~Sornette, 
"The Black-Scholes Option Pricing Problem in Mathematical Finance: 
Generalizations and Extensions for a Large Class of Stochastic Processes", 
Journal de Physique I {\bf 4}, 863 (1994).

\bibitem{Constantinides}
G.M.~Constantinides ``Stochastic Cash Management with Fixed and 
Proportional Transaction Costs", Management Science {\bf 22}, 1320-31 (1976).

\bibitem{ConstantinidesZariphopolou} 
G.M.~Constantinides and T.~Zariphopolou 
"Bounds on Derivative Prices in an Intertemporal Setting with 
Proportional Transaction Costs and Multiple Securities", Mathematical Finance 
{\bf 11}, 331-346 (2001).



\bibitem{DavisPanasZariphopolou}
M.H.~Davis, V.G.~Panas and T.~Zariphopolou, 
``European option pricing with transaction costs'', 
SIAM J. Control and Optimization {\bf 31}, 470-493 (1993).


\bibitem{DybvigRogersBack} Ph.H.~Dybvig, L.C.G.~Rogers and K.~Back,
``Portfolio Turnpikes'', {\it The Review of Financial Studies}
{\bf 12} (1999), 165-195.




\bibitem{EdhirisingheNaikUppal} 
C. Edirisinghe, V.~Naik and R.~Uppal, 
``Optimal replication of options with transaction costs and 
trading restrictions'', Journal of Financial and Quantitative Analysis 
{\bf 28} 117-138 (1993).

\bibitem{Farmer} J.D.~Farmer
``Market force, ecology and evolution'',
{\it Santa Fe Institute series Research in Economics} {\bf 98-12-117e}
(1998) and\\ 
{\tt http://www.santafe.edu/sfi/publications/Working-Papers}.


\bibitem{FlemingMeteSoner} W.H.~Fleming and H.~Mete~Soner 
{\it Controlled Markov Processes and Viscosity Solutions}, 
(Springer-Verlag, Berlin 1992).

\bibitem{Frisch} U.~Frisch
{\it Turbulence: The legacy of A. N. Kolmogorov}
(Cambridge University Press, 1995)


\bibitem{HakansonZiemba} N.~Hakanson and W.~Ziemba,
``Capital Growth Theory'',  
in {\it Handbooks in OR \& MS, Vol.9},
eds. R. Jarrow et al. (Elsevier Science, 1995).

\bibitem{KaratzasShreve}
I.~Karatzas and S.~Shreve
{\it Methods of Mathematical Finance} 
(Springer-Verlag, 1998)

\bibitem{Kelly} J.L.~Kelly~Jr.,
 ``A new interpretation of the Information Rate'',
{\it Bell Syst. Tech. J.} {\bf 35}, 917 (1956).



\bibitem{Leland}
H.E.~Leland ``Option pricing and replication with transaction costs'', 
J. Finance {\bf 40}, 1283-1301 (1985).  



\bibitem{Merton68} R.C.~Merton, "Lifetime Portfolio Selection under Uncertainty: 
The Continuous-Time Case", Review of Economics and Statistics {\bf 51}, 247-257  
(1969).

\bibitem{Merton} R.C.~Merton, ``Consumption and Portfolio Rules in
a Continuous-Time Model", Journal Of Economic Theory {\bf 3}, 373-413 (1971).


\bibitem{MG1}  P.~Muratore-Ginanneschi, 
``Models of passive and reactive tracer motions: an application of Ito calculus'',
J. Phys. A: Math. Gen. 30 (1997), L519-L523 and cond-mat/9610166. 




\end{thebibliography}
\end{document}